# Local quantum mechanics with finite Planck mass

M. Kozlowski[1], J. Marciak – Kozlowska[2]
M. Pelc[3]

[1]Corresponding author
[2] Institute of Electron Technology, Al. Lotnikow 32/46, 02 – 668, Warsaw, Poland
[3] Physics Institute Maria-Curie Sklodowska University Lublin, Poland

Abstract

In this paper the motion of quantum particles with initial mass *m* is investigated. The QM equation is formulated and solved. It is shown that the wave function contains the component which is depended on the gravitation fine structure constant

1. Introduction

Classically, when the inertial mass $m_i$ and the gravitational mass $m_g$ are equated the mass drops out of Newton's equation of motion, implying that particles of different mass with the same initial condition follows the same trajectories. But in Schrödinger's equation the masses do not cancel. For example in a uniform gravitational field [1]

$$i\hbar \frac{\partial \Psi}{\partial t} = -\frac{\hbar^2}{2m_i}\frac{\partial^2 \Psi}{\partial x^2} + m_g g x \Psi$$

implying mass dependent difference in motion.

In this paper we investigate the motion of particle with inertial mass $m_i$ in the potential field $V$. As was shown in monograph [2] the general QM equation has the form

$$i\hbar \frac{\partial \Psi}{\partial t} = V\Psi - \frac{\hbar^2}{2m}\nabla^2 \Psi - 2\tau\hbar \frac{\partial^2 \Psi}{\partial t^2}$$

where the term

$$2\tau\hbar \frac{\partial^2 \Psi}{\partial t^2}, \quad \tau = \frac{\hbar}{m_i c^2}$$

describes the memory of the particle with mass $m_i$. Above equation for the wave function $\Psi$ is the local equation with finite invariant speed, $c$ which equals the light speed in the vacuum.

2. The model equation

In this paper the local Schrödinger equation with finite Planck mass is obtained:

$$i\hbar \frac{\partial \Psi}{\partial t} = V\Psi - \frac{\hbar^2}{2m}\nabla^2 \Psi - \tau\hbar \frac{\partial^2 \Psi}{\partial t^2} \tag{1}$$

The new relaxation term (memory term)

$$\tau\hbar \frac{\partial^2 \Psi}{\partial t^2} \tag{2}$$

describes the interaction of the particle with mass $m$ with space-time.

The relaxation time $\tau$ can be calculated as

$$\tau^{-1} = \left(\tau_{e-p}^{-1} + ... + \tau_{Planck}^{-1}\right) \tag{3}$$

where $\tau_{e-p} \sim 10^{-17}$ s denotes the scattering of the particle $m$ on the electron – positron virtual pair, $\tau_{Planck} \approx 10^{-43}$ s

$$\tau_{Planck} = \frac{\hbar}{M_p c^2} \tag{4}$$

where $M_p$ is Planck mass.

Considering Eqs (1-4) Eq. (1) can be written as:

$$i\hbar \frac{\partial \Psi}{\partial t} = -\frac{\hbar^2}{2m}\nabla^2\Psi + V\Psi - \frac{\hbar^2}{2M_p}\nabla^2\Psi + \frac{\hbar^2}{2M_p}\left(\nabla^2\Psi - \frac{1}{c^2}\frac{\partial^2\Psi}{\partial t^2}\right) \qquad (5)$$

As can be seen from Eq (5) for $M_p \to \infty$ one obtains non-local Schrödinger equation

$$i\hbar \frac{\partial}{\partial t}\Psi = -\frac{\hbar^2}{2m}\nabla^2\Psi + V\Psi \qquad (6)$$

From equation (5) can be concluded that Schrödinger QM is valid for particles with $m \ll M_p$. The last term

$$\frac{\hbar^2}{2M_p}\left(\nabla^2\Psi - \frac{1}{c^2}\frac{\partial^2\Psi}{\partial t^2}\right) \qquad (7)$$

when is equal zero

$$\nabla^2\Psi - \frac{1}{c^2}\frac{\partial^2\Psi}{\partial t^2} = 0 \qquad (8)$$

describes the pilot wave equation. It is interesting to observe that the pilot wave equation is independent of mass of the particles.

Let us look for the solution of the Eq. (5), $V=0$, in the form (for 1D)

$$\Psi = \Psi(x - ct) \qquad (9)$$

For $\tau \neq 0$, i.e. for finite Planck mass we obtain:

$$\Psi(x - ct) = \exp\left(\frac{2\mu ic}{\hbar}(x - ct)\right) \qquad (10)$$

where the reduced $\mu$ mass equals

$$\mu = \frac{m_i M_p}{m + M_p} \qquad (11)$$

For $m \ll M_p$, i.e. for all elementary particles one obtains

$$\mu = m_i \qquad (12)$$

and formula (2) describes the wave function for free Schrödinger particles

$$\Psi(x - ct) = \exp\left(\frac{2mic}{\hbar}(x - ct)\right) \qquad (13)$$

For $m \gg M_p$, $\mu = M_p$

$$\Psi(x - ct) = \exp\left(\frac{2M_p ic}{\hbar}(x - ct)\right) \qquad (14)$$

From formula (6) we conclude that $\Psi(x - ct)$ is independent of $m$ of particle, $m$.

In the case $m < M_p$ from formulae (2) and (3) one obtains

$$\mu = m\left(1 - \frac{m}{M_p}\right)$$

$$\Psi(x - ct) = \exp\left(\frac{2imc}{\hbar}(x - ct)\right)\exp\left(-i\frac{m}{M_p}\left(\frac{2mc}{\hbar}x - \frac{2mc^2}{\hbar}t\right)\right)$$

(15)

In formula (7) we put

$$k = \frac{2m_i c}{\hbar}$$

$$\omega = \frac{2m_i c^2}{\hbar}$$

(16)

and obtain

$$\Psi(x - ct) = e^{i(kx - \omega t)} e^{-i\frac{m}{M_p}(kx - \omega t)}$$

(17)

As can concluded from formula (17) the second term depends on the gravity

$$\exp\left[-i\frac{m_i}{M_p}(kx - \omega t)\right] = \exp\left[-i\left(\frac{m_i^2 G}{\hbar c}\right)^{1/2}(kx - \omega t)\right]$$

(18)

where $G$ is the Newton gravity constant.

It is interesting to observe that the new constant, $\alpha_G$,

$$\alpha_G = \frac{m_i^2 G}{\hbar c}$$

(19)

is the gravitational fine structure constant. For $m_i = m_N$ nucleon mass

$$\alpha_G = 5.9042 \cdot 10^{-39}$$

3. Conclusions

In this paper the solution of the QM equation with memory term was obtained. It is shown that for $m < M_p$ the wave function $\Psi$ contains the component dependant on the fine structure constant for gravity, $\alpha_G = \frac{m_i^2 G}{\hbar c}$.